\documentclass{ptephy_v1}

\preprintnumber{XXXX-XXXX}

\usepackage{amsmath}
\usepackage{braket}
\usepackage{graphicx}
\usepackage{color}

\begin{document}

\def\beq{\begin{equation}}
\def\eeq{\end{equation}}
\def\nn{\nonumber}
\def\mr{\mathrm}
\def\s{\sigma}
\def\f{\frac}
\def\e{\mr{e}}

\title{Quantum-to-Classical Reduction of Quantum Master Equations}

\author{Norikazu Kamiya}
\address{{Department of Physics and Astronomy, Kyoto University, Sakyo-ku, Kyoto 606-8502, Japan}
\email{kamiya@scphys.kyoto-u.ac.jp}}

\begin{abstract}
A general method of quantum-to-classical reduction of quantum dynamics is described.
The key aspect of our method is the similarity transformation of the Liouvillian, which provides 
a new perspective. In conventional studies of quantum energy transport, the rotating wave 
approximation has been frequently regarded as an inappropriate approach because it causes the
energy flow through the system to vanish. Our formulation elucidates as to why this unphysical result
occurs and provides a solution for the problem. That is, not only the density matrix but also the 
physical quantity is to be transformed. Moreover, we show that quantum 
dynamics can be ``exactly" replaced with classical equations for the calculation 
of the transport efficiency. 
\end{abstract}


\maketitle

\section{Introduction}
\label{secintro}

Understanding the time evolution of an open quantum system (for e.g., a phenomenon such as quantum 
transport) is an important issue in quantum physics. One of the most common methods
to study the problem is the quantum-master-equation approach \cite{bib1.0}. In contrast to 
a classical master equation, a quantum master equation has a coherent part (off-diagonal 
elements of the density matrix), which makes the problem difficult. For example, 
the non-commutativity due to off-diagonal elements prevents analytical calculation.
The presence of the coherent part also leads to a difficulty in numerical calculation because in many-body quantum
systems, the number of off-diagonal elements of the density matrix is considerably larger than 
the number of the diagonal elements. 
These difficulties are addressed by reducing a quantum master equation
to the corresponding classical one, at least in the two following cases. One is the situation in which the 
system we consider is under the influence of environmental decoherence 
\cite{bib1.0,bib2.0,bib2.1}. The other is the decoupling between the population 
part and the coherent part achieved by the rotating wave approximation (RWA) \cite{bib1.0}. 
The resulting classical master equations are easy to calculate and they are also intuitively
understandable.
These classical reductions are uniformly described by the projection-operator technique. 
Although these methods have succeeded in many areas of quantum physics, the classical reduction of the master equation
for quantum energy transport fails as we will describe afterward.

In the current paper, we present a new method of classical reduction that involves not only
the classical reduction of the quantum master equation but also the transformation of the physical observables.
The concept underlying our method is the similarity transformation of Liouvillians, and the formulation
explicitly indicates the necessity of the transformation of the observables.
This is a crucial difference between the conventional approaches (for e.g. the projection-operator technique) 
 and our method, which solves the difficulty of the conventional approaches.
We provide the formula for the transformation of the observables in the general form.

The RWA is one of the most popular methods to obtain a quantum master equation that is 
of the Lindblad form, which is a desirable property because it 
ensures the trace preserving property and complete positivity of the density matrix \cite{bib3.0}. 
The RWA is generally applicable as long as 
the energy levels are not degenerate and the system is weakly coupled with its environment.
However, in the study of quantum energy transport, the use of the RWA leads to 
a puzzling problem \cite{bib4.0}. The density matrix in the steady state for the master equation with the 
RWA is diagonal in the energy representation, while the internal energy current operator is off-diagonal. Consequently, 
there is no resulting energy flow for the RWA master equation. Only the calculation of the 
bath-to-system energy flux is successfully performed, for example, with
the generalized quantum master equation \cite{bib5.0,bib5.1}. 
This is highly unphysical, and it has been frequently considered that the RWA is 
inappropriate for the study of energy transport. Our method that utilizes a similarity transformation clearly 
addresses the problem: the observables should also be transformed in the RWA.

\begin{table}[t]
\begin{tabular}{|c||c|c|c|} \hline
       & Redfield equation & assumed Lindblad equation & RWA-equation\\ \hline \hline
physical picture & clear & unclear & clear\\ \hline
equilibrium solution & Gibbs state & not Gibbs state & Gibbs state\\ \hline
numerical calculation & difficult & easy & easy\\ \hline
positivity & not ensured & ensured & ensured\\ \hline
\end{tabular} 
\caption{Comparison table between the Redfield equation, the assumed 
Lindblad equation, and the RWA-equation.}
\label{compari}
\end{table}

 The solution of the problem in the RWA has much significance in the study of quantum energy transport.
 Conventionally, two approaches for the problem of quantum energy transport have been proposed. One is to use 
 the Redfield equation which is derived from the total Hamiltonian including the reservoir and the interaction
 Hamiltonian \cite{bib9.0,bib9.2,bib9.3,bib9.4,bib9.5}. 
The Redfield equation is unfortunately not of the Lindblad form and does not ensure the positivity of 
 the density matrix. Moreover, it is difficult to compute the Redfield equation.   
 If we take the RWA in the Redfield equation, we obtain the classical equation that is of the Lindblad form.
 However, the RWA-equation has the problem as mentioned above and has not been much used. 
 The other approach is to start with the Lindblad equation without derivation from the total Hamiltonian \cite{doc6,doc7,doc8,doc9,doc10,bib9.1}.
 The dissipator in this case consists of local operators of the edges of the system.  Although the assumed Lindblad equation
 is easy to calculate compared to the Redfield equation, it lacks its physical picture and the stationary solution of the master
equation is not the canonical equilibrium distribution (Gibbs state) \cite{doc6}.
In fact, the assumed Lindblad equation
 can be derived from the quantum repeated interaction model (QRIM), which is a model to represent the laser-beam-like interaction 
 and does not conserve the total energy \cite{QRIM1,QRIM2}. The QRIM is thus inappropriate as a model to study the energy transport.
 The two conventional approaches have these defects respectively. In contrast, the RWA has a clear physical picture, and at the same time,  
 it ensures the positivity of the density matrix (Table \ref{compari}).

As another advantage of the idea of the similarity transformation of Liouvillans, 
We show that the quantum-to-classical reduction {\it rigorously} holds for the calculation of a quantity
such as transport efficiency. The classical picture is intuitive, and it aids in understanding environment-assisted 
quantum transport phenomena \cite{bib6.0,bib6.1,bib6.2,bib6.3}. 

The rest of this paper is organized as follows.
In section \ref{secmethod}, we present a general method of reducing a quantum Markovian master equation to
the corresponding classical one by means of utilizing a similarity transformation. In section \ref{secexample}, we apply the method to
a system under the influence of decoherence, and we confirm the validity of our method. In section \ref{secRWA}, 
we explain why the expectation value of the energy current vanishes in the steady state of the RWA master
equation, and we show that 
the transformation of observables by our method recovers the consistency. Another technique of 
quantum-to-classical reduction is presented in section \ref{secexact}. The summary of the study is provided in section \ref{secsum}. 
Throughout the paper, we set the Planck constant to unity.
    
\section{Reduction to Classical Dynamics}
\label{secmethod}

In this section, we present a method to extract the effective classical Liouvillian for
a quantum system. Our strategy is to split the eigenvalue problem of the Liouvillian
into the population part and the coherent part in a certain basis. We show that the splitting procedure 
can be carried out by a similarity transformation.

First, we define the inner product of operators $(A,B)$ as
\beq
(A,B) = \mr{Tr}[A^{\dagger}B]
\eeq
where $A$ and $B$ denote arbitrary operators. With the definition, we represent the quantum Markovian master 
equation in the block matrix form: 
\beq
\label{eq2}
\f{d}{dt}\rho = {\cal L}\rho =
\begin{pmatrix}
{\cal L}_{\mr{PP}} & {\cal L}_{\mr{PC}} \\
{\cal L}_{\mr{CP}} & {\cal L}_{\mr{CC}}
\end{pmatrix}
\begin{pmatrix}
\rho_{\mr{P}}  \\
\rho_{\mr{C}}
\end{pmatrix}.
\eeq 
Here, $\rho_{\mr{P}}$ and $\rho_{\mr{C}}$ denote the diagonal and off-diagonal components, respectively, in a certain basis 
of the density matrix $\rho$. Since in the present paper, we treat superoperators as 
matrices and operators as vectors, in order to avoid confusion, we call the subspace spanned by 
the diagonal components of operators as ``P-space" and the remnant space ``C-space." The dynamics is 
called classical if it is closed in P-space. 

We assume that the minimum value of the diagonal components of the superoperator ${\cal L}_{\mr{CC}}$ (say $M$)
is much larger than the maximum of the other components of ${\cal L}$ (say $m$).    
This is the condition of application for the method in this study. 
For example, if the energy levels of the system are not degenerate and the unitary part of the Liouvillian is large compared 
to other parts, the diagonal components of ${\cal L}_{\mr{CC}}$ in the energy representation have large values, 
which is also the condition of application for the RWA. 
In this case, the minimum of the energy level spacings corresponds to $M$. 
In the following, without loss of generality, we set $M=\Gamma$ and $m=1$ for simplicity.

The solution of the quantum master equation \eqref{eq2} is related to the eigenvalue problem
\beq{}
\label{ep}
\begin{pmatrix}
{\cal L}_{\mr{PP}} & {\cal L}_{\mr{PC}} \\
{\cal L}_{\mr{CP}} & {\cal L}_{\mr{CC}}
\end{pmatrix}
\begin{pmatrix}
\rho_{\mr{P}}^{E}  \\
\rho_{\mr{C}}^{E}
\end{pmatrix}
=E
\begin{pmatrix}
\rho_{\mr{P}}^{E}  \\
\rho_{\mr{C}}^{E}
\end{pmatrix},
\eeq
where $\rho^{E}=(\rho_{\mr P}^{E},\rho_{\mr C}^{E})^T$ denotes
the eigenvector corresponding to the eigenvalue $E$.
The above equation can be solved formally, and we obtain the equation for $\rho_{\mr{P}}^{E}$ as
\beq{}
\label{rd}
\bigl( {\cal L}_{\mr{PP}} + {\cal L}_{\mr{PC}}\f{1}{E-{\cal L}_{\mr{CC}}} {\cal L}_{\mr{CP}} \bigr)  \rho_{\mr{P}}^{E} 
= E \rho_{\mr{P}}^{E}.
\eeq{}

The eigenvalues of ${\cal L}$ are divided into two groups.
One consists of the eigenvalues of ${\cal O} (\Gamma ^0)$ and the other those of ${\cal O} (\Gamma)$. 
The latter is related to coherence decaying or fast rotating wave dynamics 
and the time scale is fast. Therefore, the dynamics can be regarded as a classical one when we focus only
on phenomena occurring at sufficiently large time scales. For an eigenvalue $\epsilon$ that has an
${\cal O} (\Gamma^0)$ value, the following equation approximately holds:
\beq{}
\label{CD}
 \biggl ( {\cal L}_{\mr{PP}} -{\cal L}_{\mr{PC}}\f{1}{{\cal L}_{\mr{CC}}}{\cal L}_{\mr{CP}}\biggr )
 \rho_{\mr{P}}^{\epsilon} = \epsilon \rho_{\mr{P}}^{\epsilon}+{\cal O}(\Gamma^{-2}).
\eeq{}
The above effective Liouvillian is nothing but that derived by the
projection-operator technique.

The above procedure can be understood in terms of a similarity transformation of ${\cal L}$.
The eigenvalue problem expressed by Eq. \eqref{ep} is transformed by an arbitrary non-singular matrix (superoperator)
${\cal S}$ without any changes in the eigenvalue:
\beq{}
{\cal S}
\begin{pmatrix}
 {\cal L}_{\mr{PP}} & {\cal L}_{\mr{PC}} \\
{\cal L}_{\mr{CP}} & {\cal L}_{\mr{CC}}
\end{pmatrix}
{\cal S}^{-1}{\cal S}
\begin{pmatrix}
\rho_{\mr{P}} ^{E} \\
\rho_{\mr{C}}^{E}
\end{pmatrix}
=E
{\cal S}
\begin{pmatrix}
\rho_{\mr{P}}^{E}  \\
\rho_{\mr{C}}^{E}
\end{pmatrix}.
\eeq
Let us define the transformed Liouvillian $ {\cal L}' \equiv {\cal S} {\cal L} {\cal S}^{-1} $
and the new density matrix ${\rho ^{\mr{E}}}' \equiv {\cal S}\rho^{\mr{E}}$.
We note that the transformation may violate the property of positive mapping, and
in fact, there exists a transformation from a positive mapping Liouvillian to a non-positive mapping Liouvillian. 
If we choose ${\cal S}$ as
\beq{}
\label{s1}
{\cal S} = 
\begin{pmatrix}
I & 0 \\
-(E-{\cal L}_{\mr{CC}})^{-1}{\cal L}_{\mr{CP}} & I
\end{pmatrix},
\eeq
 then, the density matrix is transformed as
 \beq{}
 \label{7}
 {\rho^{\mr{E}}}' = {\cal S}\rho^{\mr{E}} = 
 \begin{pmatrix}
   \rho_{\mr{P}}^{E} \\
   0
 \end{pmatrix}.
 \eeq
This is because the equality 
\beq{}
\label{8}
\rho_{\mr{C}}^{E} = \f{1}{E-{\cal L}_{\mr{CC}}}{\cal L}_{\mr{CP}} \rho_{\mr{P}}^{E}
\eeq
holds from Eq. \eqref{ep}. 
The transformed Liouvillian ${\cal L}'$ is expressed as
\beq
\label{prime}
{\cal L}'=
\begin{pmatrix}
{\cal J}  & {\cal L}_{\mr{PC}} \\
{\cal L}_{\mr{CP}} + {\cal L}_{\mr{CC}} {\cal A}-{\cal A} {\cal J} 
 &-{\cal A} {\cal L}_{\mr{PC}} + {\cal L}_{\mr{CC}}
\end{pmatrix},
\eeq
where ${\cal A}=(E-{\cal L}_{\mr{CC}})^{-1} {\cal L}_{\mr{CP}}$
and ${\cal J} = {\cal L}_{\mr{PP}} + {\cal L}_{\mr{PC}} {\cal A}$.
Utilizing Eq. \eqref{ep}, the following equality holds:
\beq
\label{9}
({\cal L}_{\mr{CP}} + {\cal L}_{\mr{CC}} {\cal A}-{\cal A} {\cal J} )\rho_{\mr{P}}^{E} =0.
\eeq 
For an eigenvalue $\epsilon$, ${\cal S}$ approximately becomes 
\beq
\label{Sapp}
{\cal S} \simeq 
\begin{pmatrix}
I & 0 \\
{\cal L}_{\mr{CC}}^{-1} {\cal L}_{\mr{CP}} & I
\end{pmatrix},
\eeq
which does not depend on eigenvalues. All the eigenvectors of ${\cal O}(\Gamma ^0)$ are 
transformed as Eq. \eqref{7}.

Next, let us consider the time evolution of the density matrix.
A density matrix is expanded by the eigenvectors of the Liouvillian, and its time evolution can be expressed as
\beq\label{expand}
\rho (t) = \e ^{{\cal L}t} \rho(0)
= \e ^{{\cal L}t}  \bigl( \sum_{\alpha} C_{\alpha} \rho^{\alpha} + \sum_{\beta} C_{\beta}^{'} \rho^{\beta} \bigr),
\eeq
where the $\rho^{\alpha}$ values denote the eigenvectors of the Liouvillian, which correspond to the eigenvalues of ${\cal O}(\Gamma^0)$, and
the $\rho^{\beta}$ values denote the eigenvectors corresponding to the eigenvalues of ${\cal O}(\Gamma)$. $C_{\alpha}$ and $C_{\beta}^{'}$
denote the coefficients of the expansion. We can ignore the summation over $\beta$ in Eq. \eqref{expand} because the
time evolution of $\rho^{\beta}$ is fast-decaying in the case wherein the real parts of the eigenvalues are large or the
time evolution is fast-rotating if the imaginary parts of the eigenvalues are large.  
Consequently, this density matrix is transformed into
\beq
\label{rhoapp}
{\cal S} \rho (0) \simeq
\begin{pmatrix}
\rho_{\mr{P}} (0) \\
0
\end{pmatrix}.
\eeq
The Liouvillian is transformed by ${\cal S}$ in Eq. \eqref{Sapp} into
\beq{}
\label{mat}
{\cal L} ' = {\cal S}{\cal L}{\cal S}^{-1} \simeq
\begin{pmatrix}
	{\cal L}_{\mr{PP}} -{\cal L}_{\mr{PC}} {\cal L}_{\mr{CC}}^{-1}{\cal L}_{\mr{CP}} & {\cal L}_{\mr{PC}} \\
	{\cal L}_{\mr{CC}}^{-1} {\cal L}_{\mr{CP}}({\cal L}_{\mr{PP}} -{\cal L}_{\mr{PC}} {\cal L}_{\mr{CC}}^{-1}{\cal L}_{\mr{CP}}) & {\cal L}_{\mr{CC}}^{-1} {\cal L}_{\mr{CP}} {\cal L}_{\mr{PC}} + {\cal L}_{\mr{CC}}
\end{pmatrix}.
\eeq{}
We can regard the left-bottom block of ${\cal L}'$ in Eq. \eqref{mat} as zero matrix
because we ignore the $\beta$-summation part and the left-bottom block of ${\cal L}'$ does not 
affect  the dynamics by virtue of Eqs. \eqref{prime}--\eqref{Sapp}.
From Eqs. \eqref{7}, \eqref{Sapp}, and \eqref{rhoapp}, we obtain the following equation:
\beq
{\cal S}\rho(t) = \e^{{\cal L}'t} {\cal S} \rho(0)
\simeq
\begin{pmatrix}
\e^{{\cal L}_{\mr{eff}}t} \rho_{\mr{P}}(0) \\
0
\end{pmatrix},
\eeq
where we define the superoperator ${\cal L}_{\mr{eff}}$ as
\beq
\label{maineq}
{\cal L}_{\mr{eff}} = {\cal L}_{\mr{PP}} -{\cal L}_{\mr{PC}} {\cal L}_{\mr{CC}}^{-1}{\cal L}_{\mr{CP}}.
\eeq
This expression is equivalent to Eq. \eqref{CD}. 
However, Eq. \eqref{CD} is considered as a mere approximation,
whereas Eq. \eqref{maineq} means the transformation. The difference in meaning causes a significant effect
on how to calculate an expectation value of observables.

The foregoing formulation utilizing a similarity transformation provides an important perspective
on the calculation of physical quantities.
The statistical average of an arbitrary physical observable $A$ can be written as:
\beq
\mr{Tr}[A \rho(t)] = \mr{Tr}[A {\cal S}^{-1} {\cal S} \rho (t)]
 = \mr{Tr}[A {\cal S}^{-1} \e^{{\cal L}_{\mr{eff}}t} \rho_{\mr{P}}(0)].
\eeq
Thus, when we reduce the quantum master equation to the classical one, the observable $A$ also should be
transformed to $\tilde{A}$ as:
\beq
\tilde{A} = A {\cal S}^{-1} = \{{\cal S}^{-1}\}^{\dagger} A,
\eeq
where we introduce the adjoint superoperator for convenience, which is defined for an arbitrary 
superoperator ${\cal K}$ as \cite{bib1.0,bib7.0}
\beq
\mr{Tr}[A{\cal K} \rho]  = \mr{Tr}[({\cal K}^{\dagger}A) \rho].
\eeq
From Eq. \eqref{Sapp}, $\tilde{A}$ is written as
\beq
\label{ObTrans}
\tilde{A} = \bigl\{1 - {\cal P} {\cal L}^{\dagger} \{(1-{\cal P}) {\cal L}^{\dagger}(1-{\cal P})\}^{-1} \bigr\} A,
\eeq 
where the superoperator ${\cal P}$ represents the projection superoperator onto the P-space.
If $A$ has no C-space components, it is not changed by ${\cal S}$. The transformation given by Eq.
\eqref{ObTrans} is a novel and important outcome of our formulation, and it explains why the 
RWA results in problems in the study of energy transport and further explains how 
the physical consistency can be recovered (section \ref{secRWA}).

\section{Example: Quantum Dynamics under Dephasing }
\label{secexample}
In this section, we apply the method described in the previous section to a simple physical system.
We consider a single particle hopping on a one-dimensional lattice with a periodic boundary
condition, which is given by the Hamiltonian:
\beq
\label{TBH}
H = \sum_{k=1}^{N} (\ket{k}\bra{k+1} + \ket{k+1}\bra{k}),
\eeq
where the ket-vector $\ket{k}$ represents the particle being at the site $k$
and $\ket{N+1}=\ket{1}$.
We assume that the system is influenced by its environment and the dynamics is described by 
the following Lindblad equation:
\beq
\label{exam}
\f{d}{dt} \rho = i[\rho,H] + \Gamma \sum_{k=1}^N (n_k \rho n_k -\f{1}{2}\{n_k,\rho\}),
\eeq
where $n_k=\ket{k}\bra{k}$ and $\{.,.\}$ denotes the anticommutator.
The non-unitary part in Eq. \eqref{exam} is called ``pure dephasing," 
which is one of the simplest models of the environmental noise,
and it has been frequently used in studies of quantum transport efficiency 
\cite{bib6.0,bib6.1,bib6.2,bib6.3} and quantum transport in the stationary state
\cite{bib8.0,bib8.1,bib8.2,bib8.3}.

Here, we represent the Liouvillian with the basis $\{\ket{k}\bra{j}\} \ \ (k,j=1,2,..,N)$. Consequently, the 
diagonal components of the matrix ${\cal L}_{\mr{CC}}$ are $-\Gamma$. Thus, 
we can apply our method if $\Gamma \gg 1$, which results in
\beq{}
\label{22}
\f{d}{dt} \rho_P = {\cal L}_{\mr{eff}}\rho_{\mr{P}} = -{\cal L}_{\mr{PC}}{\cal L}_{\mr{CC}}^{-1} 
                                    {\cal L}_{\mr{CP}} \rho_{\mr{P}}.
\eeq 
Let us next consider the operation of ${\cal L}_{\mr{eff}}$ on $n_{k}$. The superoperator 
${\cal L}_{\mr{CP}}$ yields
\beq{}
\label{23}
{\cal L}_{\mr{CP}} n_{k} = i(\ket{k}\bra{k-1}-\ket{k-1}\bra{k})
                                          - i(\ket{k+1}\bra{k}-\ket{k}\bra{k+1}).
\eeq
Although calculating the inverse operator ${\cal L}_{\mr{CC}}^{-1}$ is difficult in general,
it is approximately given by
\beq{}
\label{24}
\begin{array}{l}
{\cal L}_{\mr{CC}}^{-1} \ket{k}\bra{k+1}
                         = -\f{1}{\Gamma} \ket{k}\bra{k+1} + {\cal O}(\f{1}{\Gamma^{2}}).
\end{array}
\eeq
Using Eqs. \eqref{22}--\eqref{24}, we obtain
\beq{}
\label{diff}
{\cal L}_{\mr{eff}} n_{k} = \f{2}{\Gamma} (n_{k-1}-2n_{k}+n_{k+1}).
\eeq
This can be rewritten in the following Lindblad form
\beq
\f{d}{dt} \rho_{\mr{P}} = \f{2}{\Gamma} \sum_{k=1}^{N} \sum_{i=L,R} (L_{k,k+1}^{(i)} \rho_{\mr{P}} L_{k,k+1}^{(i)\dagger} 
                               -\f{1}{2}\{L_{k,k+1}^{(i)\dagger} L_{k,k+1}^{(i)}. \rho_{\mr{P}}\}),
\eeq
where 
\beq
L_{k,k+1}^{(L)}=\ket{k}\bra{k+1}, \ L_{k,k+1}^{(R)}=\ket{k+1}\bra{k}.
\eeq
Thus, the population dynamics shows a diffusive behaviour that obeys Eq. \eqref{diff}.
This result agrees with those of previous works \cite{bib2.0,bib8.1,bib8.2}. 

\begin{figure*}
\begin{centering}
   \includegraphics[width=10cm]{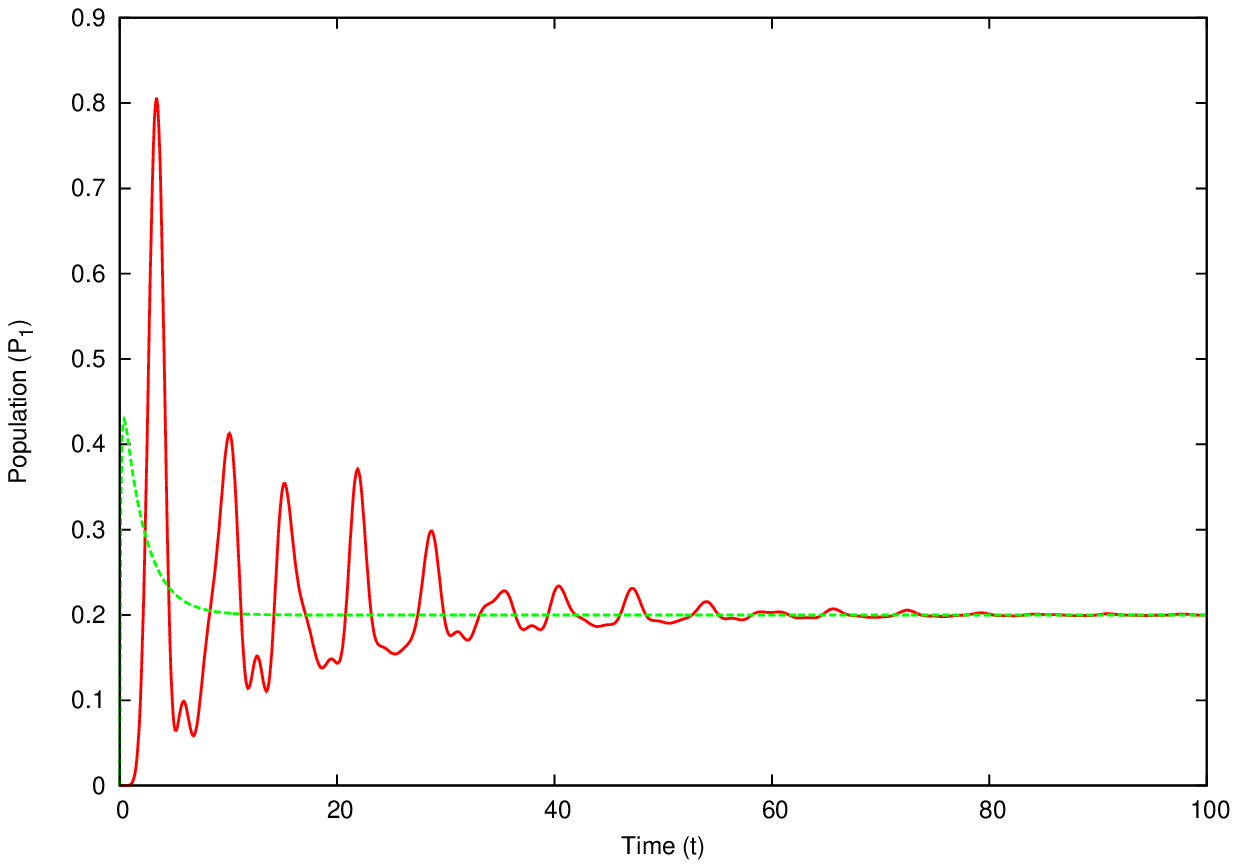}
  \includegraphics[width=10cm]{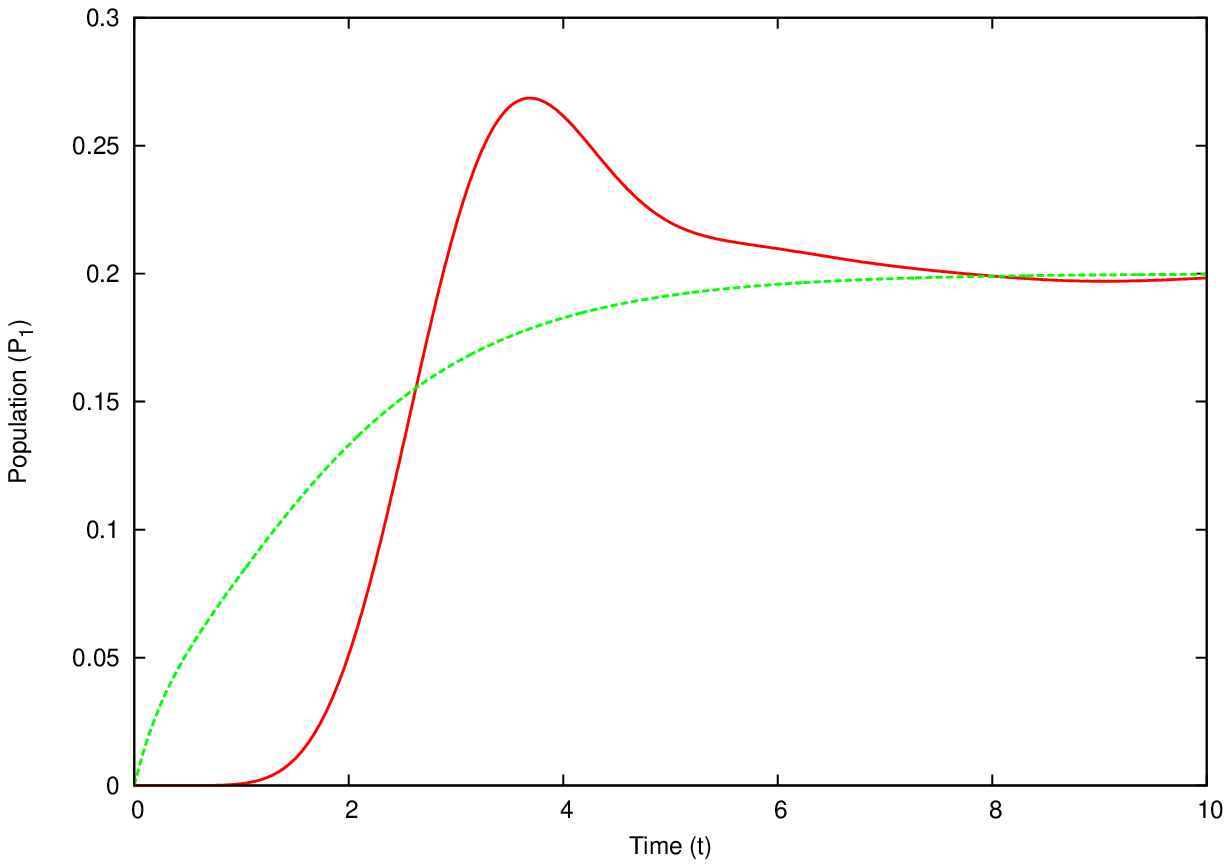}
  \includegraphics[width=10cm]{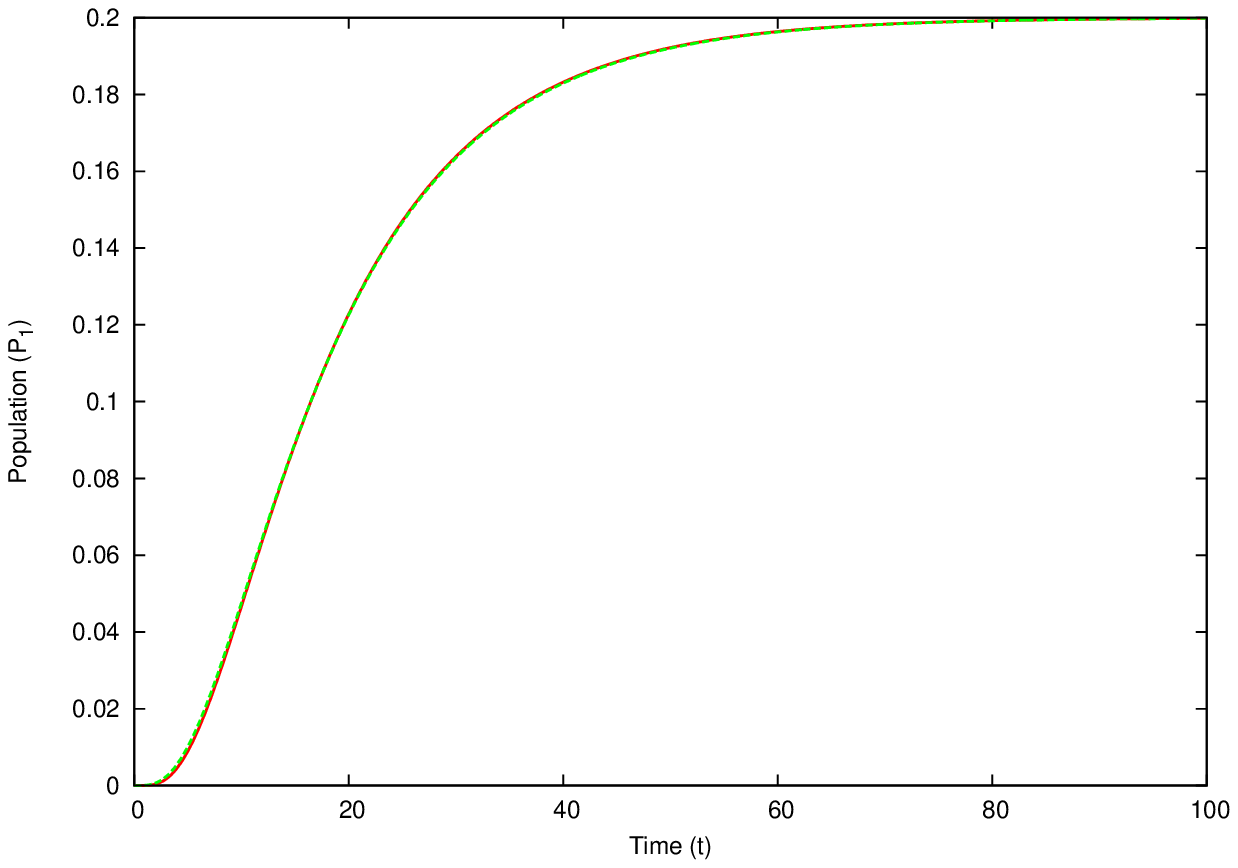}
  \caption{Time evolution of $P_1$ for three different values of dephasing rate:
$\Gamma=0.1$ (top), $\Gamma=1.0$ (middle), and $\Gamma=10$ (bottom). The red and dashed-green
lines in each panel denote results based on the original quantum dynamics \eqref{exam} and
the effective classical equation \eqref{22}, respectively.}
\label{fig1}
\end{centering}
\end{figure*}

We next compare of the original quantum dynamics with the classical 
reduction. We consider the case wherein the particle is initially at site $N$. In our study,
we numerically calculated the population at site 1, $P_1(t)=\mr{Tr}[n_1\rho(t)]$ for the system 
size $N=5$ and three different values of the dephasing rate $\Gamma$ (Fig. \ref{fig1}). In all 
cases, the population $P_1$ converges to $1/5$. However, the intermediate behaviour of the reduced
equation is different from the original quantum dynamics for $\Gamma=0.1$ and $\Gamma=1$. In contrast, 
for the case of $\Gamma=10$, the classical and the quantum time evolutions agree with each other.   
This is because quantum effects decreases as the dephasing rate increases.

In the above example, we consider the quantity $n_k$ which does not vary under the transformation \eqref{ObTrans}.
The work of the transformation \eqref{ObTrans} is shown in the following section.

\section{Energy Transport and RWA}
\label{secRWA}

In this section, we show that our formulation includes the RWA that is a standard method to study 
quantum open systems. By means of our method, we can clearly explain why the RWA gives 
unphysical results for energy transport problems. 
  
A system in contact with heat reservoirs is frequently described by the Redfield equation
\cite{bib9.0,bib9.2,bib9.3,bib9.4,bib9.5}.
It is derived from the total Hamiltonian:
\beq
H_{\mr{tot}} = H_{\mr{S}} + \lambda H_{\mr{SB}} + H_{\mr{B}},
\eeq
where $H_{\mr{S}}$, $H_{\mr{SB}}$, and $H_{\mr{B}}$ denote the system Hamiltonian, the system--bath
interaction Hamiltonian, and the bath Hamiltonian respectively. The system--bath coupling $\lambda$
is assumed to be weak. Here, we assume $H_{\mr{SB}}$ as the following:
\beq
H_{\mr{SB}} = X \otimes Y,
\eeq 
where $X$ and $Y$ denote Hermitian operators that operate on the Hilbert space of the system and that of the bath, 
respectively. Utilizing the second-order perturbation with several approximations, 
the Redfield equation is obtained in the following form:
\beq
\label{Redfield}
\f{d}{dt} \rho ={\cal L}^{(\mr{Red})} \rho = i[\rho,H_{\mr{S}}] + 
\pi \lambda^2 \sum_{i,j} \bigl ( \Gamma_{ij} \bra{\epsilon_{i}}X\ket{\epsilon_{j}}[\ket{\epsilon_i}\bra{\epsilon_j}\rho ,X] + h.c. \bigr ),
\eeq
where $\ket{\epsilon_i}$ denotes the energy eigenstate of the eigenvalue $\epsilon_i$ of $H_{\mr{S}}$ and $\Gamma_{ij}$
denote the Fourier transformations of the reservoir correlation function. The temperature of the reservoir
$\beta$ is given by the Kubo--Martin--Schwinger condition:
\beq
\Gamma_{ij} =\e^{-\beta(\epsilon_i -\epsilon_j)} \Gamma_{ji}.
\eeq 
The RWA is usually carried out by considering the interaction picture, which results in the following classical 
master equation \cite{bib1.0}:
\beq
\label{RWA}
\f{d}{dt} P_{k}(t) =
2 \pi \lambda^2 \sum_{i} \bigl( \Gamma_{ki} |X_{ik}|^{2}P_{i}(t) - \Gamma_{ik}|X_{ik}|^{2} P_{k}(t) \bigr),  
\eeq  
where $P_{k}(t)$ denotes the probability of observing the energy $\epsilon_k$ at time $t$.  

It is easily verified that Eq. \eqref{RWA} is equivalent to time evolution by 
${\cal L}^{(\mr{Red})}_{\mr{PP}}$ with the energy representation. 
The Redfield equation \eqref{Redfield} is obtained by means of 
the second-order perturbation with respect to $\lambda$. Hence, for the same level of accuracy, 
the effective classical Liouvillian \eqref{maineq} becomes 
${\cal L}_{\mr{eff}} \simeq {\cal L}^{(\mr{Red})}_{\mr{PP}}$. This indicates that the
procedure of the RWA represented by Eq. \eqref{Redfield} through Eq. \eqref{RWA} is accounted for in 
our formulation, and it also indicates that at the same time the observables should be 
transformed by Eq. \eqref{ObTrans}.

As an example, let us calculate the expectation value of the energy current 
in the steady state for the following non-equilibrium systems:
\beq
\label{EX1}
\left\{
\begin{array}{l}
{\displaystyle
\f{d}{dt} \rho = i[\rho,H] + \gamma (\s_1^{-} \rho \s_1^{+} - \f{1}{2} \{\s_1^{+} \s_{1}^{-}, \rho\}) +
\gamma (\s_2^{+} \rho \s_2^{-} - \f{1}{2} \{\s_2^{-} \s_{2}^{+}, \rho\}) }  \\
H=\ket{1}\bra{2}+\ket{2}\bra{1} + h (n_1 + n_2).
\end{array}
\right.
\eeq
where $\s_k^{+}=\ket{k}\bra{0}$, $\s_k^{-}=\ket{0}\bra{k}$, and $\ket{0}$ denotes the vacuum state.
Here we are only concerned with the single exciton space, that is, we exclude the state 
$\s_{1}^{+}\s_{2}^{+}\ket{0}$.
Although we introduce the above quantum master equation a priori here, the RWA can be performed
for weak coupling, i.e., small values of $\gamma$. The eigenvalues of the Hamiltonian are $\epsilon_0=0$, $\epsilon_1=h-1$, 
and $\epsilon_2=h+1$. The RWA can be performed if $\epsilon_{i}-\epsilon_{j} \gg \gamma\ (i\neq j)$
is satisfied, and it can be realized by assuming a suitable value of $h$. 
We define the energy current $J$ as
\beq
J = -ih (\ket{1}\bra{2}-\ket{2}\bra{1})
\eeq
As in the case of the Redfield equation, for the purpose of simplicity, we ignore the second- and
higher-order $\gamma$ terms. 

We first calculate the expectation of the energy current $\langle J \rangle$ in the 
steady state without using the RWA. The bracket $\langle \ \rangle$ denotes 
the statistical average of the observables.
The time derivative of an observable $A$ is given by
\beq
\label{adjoint}
\f{d}{dt} A = {\cal L}^{\dagger} A 
= -i[A,H] + \gamma (\s_1^{+} A \s_1^{-} - \f{1}{2} \{\s_1^{+} \s_{1}^{-}. A\})
+ \gamma (\s_2^{-} A \s_2^{+} - \f{1}{2} \{\s_2^{-} \s_{2}^{+}. A\}).
\eeq
Using this equation, we can write the time derivatives of the expectation values of observables by
\beq
\label{Eqs}
\left\{
\begin{array}{l}
{\displaystyle \f{d}{dt} \langle J \rangle = 
-2h (\langle n_1 \rangle -\langle n_2 \rangle) -\f{\gamma}{2} \langle J \rangle } \\
{\displaystyle \f{d}{dt} \langle n_1 \rangle = 
\f{1}{h} \langle J \rangle  -  \gamma\langle n_1 \rangle } \\
{\displaystyle \f{d}{dt} \langle n_2 \rangle = 
-\f{1}{h} \langle J \rangle  + \gamma\langle n_0 \rangle }.
\end{array}
\right.
\eeq
Moreover, the completeness relation 
\beq
\label{complete}
\sum_{k=0}^{2} n_k = 1
\eeq
holds. In the steady state, the left-hand sides of the equations \eqref{Eqs} vanish. Solving Eqs.  
\eqref{Eqs} and  \eqref{complete}, we obtain the expectation value of the energy current in the steady state:
\beq
\langle J \rangle \simeq \f{\gamma h}{3},
\eeq
where the higher order of $\gamma$ has been omitted.

Next, we compute the energy current $J$ with the RWA. 
The RWA results in the classical Liouvillian:
\beq
\label{RWAex}
{\cal L}_{\mr{eff}} = \f{\gamma}{2}
\begin{pmatrix}
-2 & 1 &  1 \\
 1  & - 1 & 0 \\
 1 & 0 & - 1 
\end{pmatrix},
\eeq
where the basis $\{ \ket{\epsilon_0}\bra{\epsilon_0}, \ket{\epsilon_1}\bra{\epsilon_1},
\ket{\epsilon_2}\bra{\epsilon_2} \}$ is used in this order. 
The steady-state solution of the Liouvillian \eqref{RWAex} is given as 
\beq
\label{RWASS}
\rho^{\mr{ss}} = \f{1}{3} (\ket{\epsilon_0}\bra{\epsilon_0} + \ket{\epsilon_1}\bra{\epsilon_1}
+ \ket{\epsilon_2}\bra{\epsilon_2}).
\eeq
The energy current is expressed in the energy basis as
\beq
J = ih(\ket{\epsilon_1}\bra{\epsilon_2} - \ket{\epsilon_2}\bra{\epsilon_1}).
\eeq
We note that the expression for $J$ has no P-space components.
In conventional approaches, the energy current $J$ is used without any changes, thereby 
resulting in $\mr{Tr}[J \rho^{\mr{ss}}]=0$. This brings to light the necessity of the transformation of
$J$ to $\tilde{J}$. From Eqs. \eqref{ObTrans} and \eqref{adjoint}, 
the transformed current $\tilde{J}$ is given as
\beq
\label{tilJ}
\tilde{J} \simeq  J + \f{\gamma h}{4} 
(2 \ket{\epsilon_0}\bra{\epsilon_0} + \ket{\epsilon_1}\bra{\epsilon_1}+\ket{\epsilon_2}\bra{\epsilon_2}),
\eeq
which reproduces the correct expectation value of the energy current 
\beq
\langle \tilde{J} \rangle = \mr{Tr}[\tilde{J} \rho^{\mr{ss}}]=\f{\gamma h}{3}.
\eeq

\section{Exact Replacement with Classical Dynamics}
\label{secexact}

The reduced equation \eqref{maineq} is derived by expanding the original quantum master equation with respect to
$1/\Gamma$. Therefore, it is only valid for large values of $\Gamma$. In this section, we show that the replacement 
of quantum dynamics with the classical equation can be carried out for any values of the parameters for quantities
such as transport efficiency.

Let us consider the quantum open system that is described by the Markovian quantum master equation:
\beq
\f{d}{dt}\rho = {\cal L}\rho.
\eeq 
Let us assume that the quantum master equation has a unique steady state $\rho^{\mr{ss}}$. 
Let us consider the following quantity:
\beq
\label{K}
\zeta=\int^{\infty}_{0} dt \mr{Tr}[A\e^{{\cal L}t}\rho(0)],
\eeq
where $A$ denotes a P-space observable that satisfies $\mr{Tr}[A\rho ^{\mr{ss}}]=0$. 

We first show that the time integral in Eq. \eqref{K} is related to a certain steady-state problem
from the analogy of the linear-response theory. For this purpose, we modify the Liouvillian ${\cal L}$  to ${\cal K}$ as
\beq
{\cal K} ={\cal L} + \epsilon \chi,
\eeq
where $\epsilon$ is the small parameter and the superoperator $\chi$ satisfies
\beq
\chi = {\cal P} \chi {\cal P},\ \ \chi \rho^{\mr{ss}}= \chi \rho_{\mr{P}}^{\mr{ss}} = \rho(0),
\eeq 
where $\rho_{\mr{P}}^{\mr{ss}}$ denotes the P-space components of $\rho^{\mr{ss}}$.
The steady-state solution $\eta$ of the modified Liouvillian ${\cal K}$ is expressed as
\beq
\label{sekibun}
\eta = \rho^{\mr{ss}} + \epsilon \int^{\infty}_{0} dt \e^{{\cal L}t} \chi \rho^{\mr{ss}} + {\cal O}(\epsilon ^2)
=\rho^{\mr{ss}} + \epsilon \int^{\infty}_{0} dt \e^{{\cal L}t} \rho(0) + {\cal O}(\epsilon ^2).
\eeq
Thus, $\zeta$ can be represented as
\beq
\label{kyokugen}
\zeta= \lim_{\epsilon \to 0}  \epsilon^{-1} \mr{Tr}[A\eta] = \lim_{\epsilon \to 0}  \epsilon^{-1} \mr{Tr}[A\eta _{\mr{P}}],
\eeq
where $\eta_{\mr{P}}$ represents the P-space components of $\eta$.

The steady-state problem is expressed as
\beq
\begin{pmatrix}
{\cal L}_{\mr{PP}} + \epsilon \chi & {\cal L}_{\mr{PC}} \\
{\cal L}_{\mr{CP}}  &  {\cal L}_{\mr{CC}}
\end{pmatrix}
\begin{pmatrix}
\eta_{\mr{P}} \\
\eta_{\mr{C}}
\end{pmatrix}
= 0.
\eeq
Let us transform the above equation by the superoperator ${\cal S}$
\beq{}
\label{S}
{\cal S} = 
\begin{pmatrix}
I & 0 \\
{\cal L}_{\mr{CC}}^{-1} {\cal L}_{\mr{CP}} & I
\end{pmatrix}.
\eeq
From Eqs. \eqref{s1} and \eqref{7}, we obtain the following equation:
\beq
{\cal S} {\cal L} {\cal S}^{-1} {\cal S}\eta =
\begin{pmatrix}
{\cal L}_{\mr{eff}} + \epsilon \chi & {\cal L}_{\mr{PC}} \\
{\cal L}_{\mr{CC}}^{-1} {\cal L}_{\mr{CP}} ({\cal L}_{\mr{eff}}+\epsilon \chi)  
& {\cal L}_{\mr{CC}}^{-1} {\cal L}_{\mr{CP}} {\cal L}_{\mr{PC}} + {\cal L}_{\mr{CC}}
\end{pmatrix}
\begin{pmatrix}
\eta_{\mr{P}} \\
0
\end{pmatrix}
= 0.
\eeq
Thus, $\eta_{\mr{P}}$ is the steady-state solution of the superoperator ${\cal L}_{\mr{eff}} + \epsilon \chi$,
and it can be expressed as
\beq
\label{sekibun2}
\eta_{P} = \rho_{\mr{P}}^{\mr{ss}} + \epsilon \int^{\infty}_{0} dt \e^{{\cal L}_{\mr{eff}}t} \rho(0) + {\cal O}(\epsilon ^2).
\eeq
Using Eqs. \eqref{sekibun}, \eqref{kyokugen}, and \eqref{sekibun2}, we obtain the following equation:
\beq
\zeta=\int^{\infty}_{0} dt \mr{Tr}[A\e^{{\cal L}t}\rho(0)] = \int^{\infty}_{0} dt \mr{Tr}[A\e^{{\cal L}_{\mr{eff}}t}\rho(0)].
\eeq
Thus, the time evolution of the quantum system is fully replaced by 
population dynamics. 
\begin{figure}
\centering
\includegraphics[width=10cm]{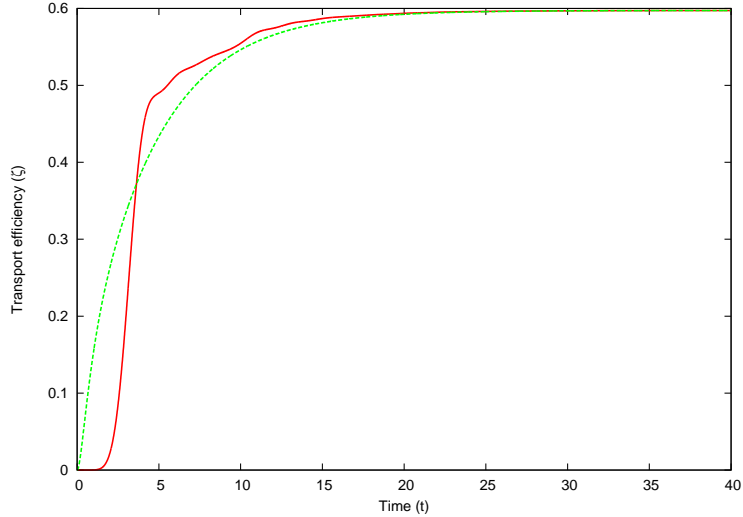}
\caption{Numerical calculation of $\zeta(t)$ based on the original quantum dynamics (red line) and the classical 
dynamics (green dashed-line). The original quantum evolution asymptotically coincides 
with the classical replacement in the long-time regime. }
\label{fig2}
\end{figure}

To validate the above argument, we numerically calculate the quantum transport efficiency for the system 
given by the following equation:
\beq
\label{QE}
\f{d}{dt} \rho = {\cal L}\rho = i[\rho,H] + {\cal L}_{\mr{deph}} \rho +{\cal L}_{\mr{diss}}\rho+{\cal L}_{\mr{trap}} \rho,
\eeq
where the Hamiltonian $H$ is the tight-binding model given by Eq. \eqref{TBH}, and the Lindblad superoperators are given by
\beq
\label{Lin3}
\left\{
\begin{array}{l}

{\cal L}_{\mr{deph}}= \gamma {\displaystyle \sum_{k=1}^{N} (n_k \rho n_k -\f{1}{2}\{n_k,\rho \})}   \\

{\cal L}_{\mr{diss}}={\displaystyle \mu \sum_{k=1}^{N} ( \s_{k}^{-} \rho \s_{k}^{+} -\f{1}{2}\{ \s_{k}^{+}\s_{k}^{-},\rho\})}    \\

{\cal L}_{\mr{trap}}= \kappa ( \s_{1}^{-} \rho \s_{1}^{+} -\f{1}{2}\{ \s_{1}^{+}\s_{1}^{-},\rho\}). \\
\end{array}
\right.
\eeq
The raising and lowering operators at the $k$-th site are denoted by $\s_{k}^{+}$ and $\s_{k}^{-}$, respectively. 
Let us suppose that the particle is at the $N$-th site initially. The transport efficiency is defined by
how often the particle is trapped at the first site during the time interval $t$, which efficiency is expressed by
\beq
\zeta(t) = \kappa \int^{t}_{0} ds \mr{Tr}[n_{1} \e^{{\cal L}s}\rho(0)],
\eeq
where $\zeta(t)$ represents the transport efficiency at time $t$. 
We numerically calculate the time evolution
of $\zeta(t)$ with parameters $N=5$, $\kappa=1$, $\mu=0.1$, and $\gamma=0.1$ (Fig. \ref{fig2}).  
The classical time evolution based on the Liouvillian ${\cal L}_{\mr{eff}}$ does not coincide with the original quantum 
time evolution in the short-time regime; however, the time evolutions converge with the same value in the long-time
regime.　  

We note that the superoperator ${\cal L}_{\mr{eff}}$ does not ensure the positivity of the density matrix
in general. Nevertheless, in the case of the transport efficiency problem, the time evolution can be intuitively
interpreted. This is because the superoperator ${\cal L}_{\mr{eff}}$ is trace-preserving. 
Therefore, the particle flow can be defined. The only difference with respect to the general classical 
picture is that negative values of population
can be obtained.

\section{Conclusions} 
\label{secsum}
We have proposed a general method to reduce a quantum master equation to a classical one by
utilizing a similarity transformation. Our formulation reveals the necessity of the 
transformation of observables.
This is the solution of the problem that the energy flow through the system vanishes
in the RWA scheme. We have also shown that the exact replacement with classical
dynamics is possible for the calculation of a quantity such as the transport efficiency. Our method
facilitates an understanding of several mechanisms of environment-assisted quantum transport 
in the unified picture \cite{prepare}.  

The introduction of a similarity transformation is also observed in 
the study of the non-relativistic reduction of the Dirac equation with electromagnetic fields
\cite{bib10.0,bib10.1}. In such a case, 
the transformation is performed on the Hamiltonian, and hence, it should be a unitary transformation. 
However, the nature of the similarity transformation in the Liouville space has not been understood
clearly. Thus, it is important to examine as to what kinds of transformations conserve the nature of
the Liouvillian that is of the Lindblad form.

The argument in this paper is general, and therefore, we expect that the results can be applied to 
a wide range of quantum physics problems.

\section*{Acknowledgments}
The author is grateful to S. Takesue for useful comments and critical reading of the manuscript. 
He also thanks one of the referees for valuable comments.


\begin{thebibliography}{50}
\bibitem{bib1.0} 
H-P. Breuer and F. Petruccione,
{\it{The Theory of Open Quantum Systems}} (Oxford University Press, Oxford, 2002)
\bibitem{bib2.0} 
S. Hoyer, M. Sarovar, and K. B. Whaley,
New J. Phys. \textbf{12}, 065041 (2010)
\bibitem{bib2.1} 
H. Haken and P. Reineker,
Z. Phys. \textbf{249}, 253 (1972)
\bibitem{bib3.0} 
G. Lindblad,
Commun. Math. Phys. \textbf{48}, 119 (1976)
\bibitem{bib4.0}
H. Wichterich, M. J. Henrich, H-P. Breuer, J. Gemmer, and M. Michel,
Phys. Rev. E \textbf{76}, 031115 (2007)
\bibitem{bib5.0} 
M. Esposito, U. Harbola, and S. Mukamel,
Rev. Mod. Phys. \textbf{81}, 1665 (2009)
\bibitem{bib5.1} 
T. Yuge, T. Sagawa, A. Sugita, and H. Hayakawa,
Phys. Rev. B \textbf{86}, 235308 (2012)
\bibitem{bib9.0}
K. Saito, S. Takesue, and S. Miyashita,
Phys. Rev. E \textbf{61}, 2397 (2000)
\bibitem{bib9.2}
T. Prosen and B. \v{Z}unkovi\v{c},
New J. Phys. \textbf{12}, 025016 (2010)
\bibitem{bib9.3} 
J. Wu and M. Berciu,
Phys. Rev. E \textbf{83}, 214416 (2011)
\bibitem{bib9.4}
M. Michel and O. Hess, Phys. Rev. B. \textbf{77}, 104303 (2008)
\bibitem{bib9.5}
K. Saito, Europhys Lett. \textbf{61}, 34 (2003)
\bibitem{doc6} 
M. Znidaric, T. Prosen, G. Benetti, G. casati, and D. Rossini, Phys. Rev. E. \textbf{81}, 051135 (2010)
\bibitem{doc7}
T. Prosen, New J. Phys. \textbf{10}, 043026 (2008)
\bibitem{doc8}
R. Steinigeweg, M. Ogiewa, and J. Gemmer, EPL \textbf{87}, 10002 (2009) 
\bibitem{doc9}
K. Sun, C. Wang, and Q. Chen,  EPL \textbf{92}, 24002 (2010)
\bibitem{doc10}
W. Li and P. Tong, Phys. Rev. E \textbf{83}, 031128 (2011)
\bibitem{bib9.1} 
Y. Yan, C.-Q. Wu, G. Casati, T. Prosen, and B. Li, 
Phys. Rev. B \textbf{77}, 172411 (2008)
\bibitem{QRIM1}
S. Attal and Y. Pautrar, Ann. Henri Poincar{\'e} \textbf{7}, 59 (2006)
\bibitem{QRIM2}
A. Dhahri, J. Phys. A \textbf{41}, 275305 (2008)
\bibitem{bib6.0}
M. B. Plenio and S. F. Huelga,
New J. Phys. \textbf{10}, 113019 (2008)
\bibitem{bib6.1}
P. Rebentrost, M. Mohseni, I. Kassal, S. Lloyd, and A. Aspuru-Guzik,
New J. Phys. \textbf{11}, 033003 (2009)
\bibitem{bib6.2}
I. Kassal and A. Aspuru-Guzik,
New J. Phys. \textbf{14}, 053041 (2012)
\bibitem{bib6.3}
K. M. Pelzer, A. F. Fidler, G. B. Griffin, S. K. Gray, and G. S. Engel,
New J. Phys. \textbf{15}, 095019 (2013)
\bibitem{bib7.0}
N. Kamiya and S. Takesue,
J. Phys. Soc. Jpn. \textbf{82}, 114002 (2013)
\bibitem{bib8.0}
M. \v{Z}nidari\v{c},
New J. Phys. \textbf{12}, 043001 (2010)
\bibitem{bib8.1}
M. \v{Z}nidari\v{c}, 
J. Stat. Mech. \textbf{2010}, L05002 (2010)
\bibitem{bib8.2}
D. Manzano, M. Tiersch, A. Asadian, and H. J. Briegel,
Phys. Rev. E \textbf{86}, 061118 (2012)
\bibitem{bib8.3}
J. J. Mendoza-Arenas, T. Grujic, D. Jaksch, and S. R. Clark,
Phys. Rev. E \textbf{87}, 235130 (2013)
\bibitem{prepare}
in preparation.
\bibitem{bib10.0}
L. L. Foldy and S. A. Wouthuysen,
Phys. Rev. \textbf{78}, 29 (1950)
\bibitem{bib10.1}
S. Okubo,
Prog. Theor. Phys. \textbf{12}, 603 (1954)
\end{thebibliography}
\end{document}